\documentclass{emulateapj}

\received{}
\accepted{}
\slugcomment{Revised Manuscript, 23 Apr. 2012}
\shorttitle{{\it Chandra} Observations of Galaxy Zoo Mergers}
\shortauthors{Teng et al.}

\begin{document}

\title{{\it Chandra} Observations of Galaxy Zoo Mergers: Frequency of Binary Active Nuclei in Massive Mergers}

\author{Stacy H. Teng~\altaffilmark{1, 2, 11}, Kevin Schawinski~\altaffilmark{3, 4, 12}, C. Megan Urry~\altaffilmark{3, 4, 5}, Dan W. Darg~\altaffilmark{6}, Sugata Kaviraj~\altaffilmark{6}, Kyuseok Oh~\altaffilmark{7}, Erin W. Bonning~\altaffilmark{3, 4}, Carolin N. Cardamone~\altaffilmark{8}, William C. Keel~\altaffilmark{9}, Chris~J.~Lintott~\altaffilmark{6}, Brooke D. Simmons~\altaffilmark{4, 5}, \& Ezequiel Treister~\altaffilmark{10}}

\altaffiltext{1}{Observational Cosmology Laboratory, NASA/GSFC, Greenbelt, MD 20771, USA; stacy.h.teng@nasa.gov}
\altaffiltext{2}{Department of Astronomy, University of Maryland, College Park, MD 20742, USA}
\altaffiltext{3}{Department of Physics, Yale University, New Haven, CT 06511, USA}
\altaffiltext{4}{Yale Center for Astronomy and Astrophysics, Yale University, P.O. Box 208121, New Haven, CT 06520, USA}
\altaffiltext{5}{Department of Astronomy, Yale University, New Haven, CT 06511, USA}
\altaffiltext{6}{Department of Physics, University of Oxford, Keble Road, Oxford, OX1 3RH, UK}
\altaffiltext{7}{Department of Astronomy, Yonsei University, Seoul 120-749, Republic of Korea}
\altaffiltext{8}{The Harriet W. Sheridan Center for Teaching and Learning, Brown University, P.O. Box 1912, Providence, RI, 02912, USA}
\altaffiltext{9}{Department of Physics \& Astronomy, 206 Gallalee Hall, 514 University Blvd., University of Alabama, Tuscaloosa, AL 35487-034, USA}
\altaffiltext{10}{Departamento de Astronom\`{i}a, Universidad de Concepci\'{o}n, Casilla 160-C, Concepci\'{o}n, Chile }
\altaffiltext{11}{NASA Postdoctoral Program Fellow}
\altaffiltext{12}{Einstein Fellow}

\begin{abstract}

We present the results from a {\it Chandra} pilot study of 12 massive galaxy mergers selected from Galaxy Zoo.  The sample includes major mergers down to a host galaxy mass of 10$^{11}$~$M_\odot$ that already have optical AGN signatures in at least one of the progenitors.  We find that the coincidences of optically selected active nuclei with mildly obscured ($N_H \lesssim 1.1 \times 10^{22}$~cm$^{-2}$) X-ray nuclei are relatively common (8/12), but the detections are too faint ($< 40$ counts per nucleus; $f_{2-10~keV} \lesssim 1.2 \times 10^{-13}$~erg~s$^{-1}$~cm$^{-2}$) to reliably separate starburst and nuclear activity as the origin of the X-ray emission.  Only one merger is found to have confirmed binary X-ray nuclei, though the X-ray emission from its southern nucleus could be due solely to star formation.  Thus, the occurrences of binary AGN in these mergers are rare (0--8\%), unless most merger-induced active nuclei are very heavily obscured or Compton thick.  

\end{abstract}


\section{Introduction}
\label{sec:intro}




Major mergers are a key component of current models for galaxy formation in a $\Lambda$CDM Universe. 
Mergers can disrupt the star-forming gas and stellar disks of the progenitors, trigger a powerful burst of star formation, and reshape the remaining stellar content into a bulge. Perhaps with a small time delay, the supermassive black holes may feed on gas from the destabilized or destroyed disk, injecting energy in the form of radiation or kinetic outflows that sweep the remnant clear of dust and gas. First proposed by \citet{sanders88}, this picture directly links the triggering of active galactic nuclei (AGNs) phases to the destructive potential induced by a merger. Recent semi-analytic models and hydrodynamic simulations have adopted this scenario to explain the fueling of AGNs and the red spheroidal remnants that are difficult to reproduce without some kind of ``AGN feedback'' \citep{springel, dimatteo, hopkins06, hopkins08, somerville}.

In principle, major mergers carry with them two black holes, both of which may be accreting and be visible as distinct AGN during a phase of abundant gas availability that a major, gas-rich merger represents. Yet the evidence associating AGN phases with major mergers remains contested \citep{derobertis, malkan, schmitt, pierce, georgakakis, gabor, schawinski11}.  Large optical surveys using Sloan Digital Sky Survey (SDSS) data have found $\sim$3.6\% of spectroscopically confirmed AGNs are in closed binaries \citep[$\sim$5--100~kpc separation;][]{liu11}.  The DEEP2 survey also found that binary AGN exist in $\sim$2.2\% (2/91) of red galaxies with type 2 Seyfert optical spectra \citep{deep2a, deep2b}.  However, optical surveys can easily miss obscured AGNs especially in merger systems where the gas is driven toward the center through dissipation \citep[e.g.,][]{hopkins08}.  X-ray surveys are needed to identify the more highly obscured systems ($N_H \gtrsim 10^{20}$~cm$^{-2}$).   We know of only a small number of binary AGN resolved directly using X-ray observations \citep[e.g.,][]{komossa, guainazzi, hudson, bianchi, foreman, comerford, fabbiano}.  The intrinsic frequency of binary AGN phases has not been observationally constrained, as the separation of individual X-ray sources is not possible in high-redshift sources and there has been no systematic search for such systems in known mergers.  A study of the host galaxies of 185 nearby ($z \lesssim 0.05$) BAT AGNs by \citet{koss11} found that these hard X-ray selected AGNs are preferentially found in massive galaxies with large bulge-to-disk ratios and large supermassive black holes.  This may imply that the frequency of binary AGNs is higher in massive mergers.

In order to quantify the intrinsic frequency of double AGNs in the local universe, we embarked on a study of the presence of binary AGNs and their dependence on the mass of the host galaxies.  The results from the present survey represent a pilot effort as the sample is comprised of only the most massive galaxies in the Galaxy Zoo merger sample.  Thus, this paper aims to quantify the intrinsic frequency of double AGNs in the mass limit down to $\sim$10$^{11}$~$M_\odot$ using a study of 12 merging galaxies with the {\it Chandra} X-ray observatory. 
Throughout this paper, we adopt $H_0 = 71$~km~s$^{-1}$~Mpc$^{-1}$, $\Omega_M = 0.3$, and $\Omega_\Lambda = 0.7$. 

\section{Sample Selection} 
\label{sec:sample}

\begin{figure}
\centering
\includegraphics[width=3in, angle=90]{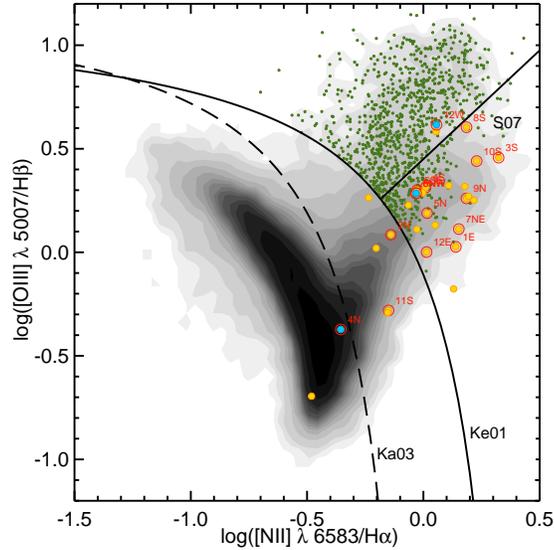}
\caption{ Emission line diagnostic diagram used for the selection of the {\it Chandra} Galaxy Zoo sample.  The grey scale represents the complete Galaxy Zoo sample and the green points are the Galaxy Zoo AGN selected using narrow line diagnostics from \citet{schawinski10}.  The dashed curve shows the empirical separation between purely star-forming galaxies and the composite region of the diagram as determined by \citet[][Ka03]{ka03}.  The solid curve is the theoretical extreme starburst line of \citet[][ke01]{ke01} beyond which the dominant source of ionization must be due to something other than star formation.  The straight line demarcates the empirical AGN-LINER separation in \citet[][S07]{schawinski07}.  The yellow points are the Galaxy Zoo mergers that meet the criteria of mergers having mass ratios of at least 3:1 and having one of the progenitors with a mass $> 10^{11} M_\odot$ with significant emission line detections; most are LINERs.  The blue points are the secondary nuclei with SDSS optical spectra.  The {\it Chandra} observed nuclei (both primary and secondary) are circled in red with their GZ identifier from Table~\ref{tab:sample} labeled.  Our {\it Chandra} sample covers the full range of [O~III]/H$\beta$ and [N~II]/H$\alpha$ emission line ratios for composite and LINER objects.  They are also representative of the merger distribution in the composite and LINER areas of this Baldwin-Phillips-Terlevich (BPT) diagram \citep{bpt81}. }
\label{fig:select}
\end{figure}

\begin{figure}
\centering
\includegraphics[scale=0.4]{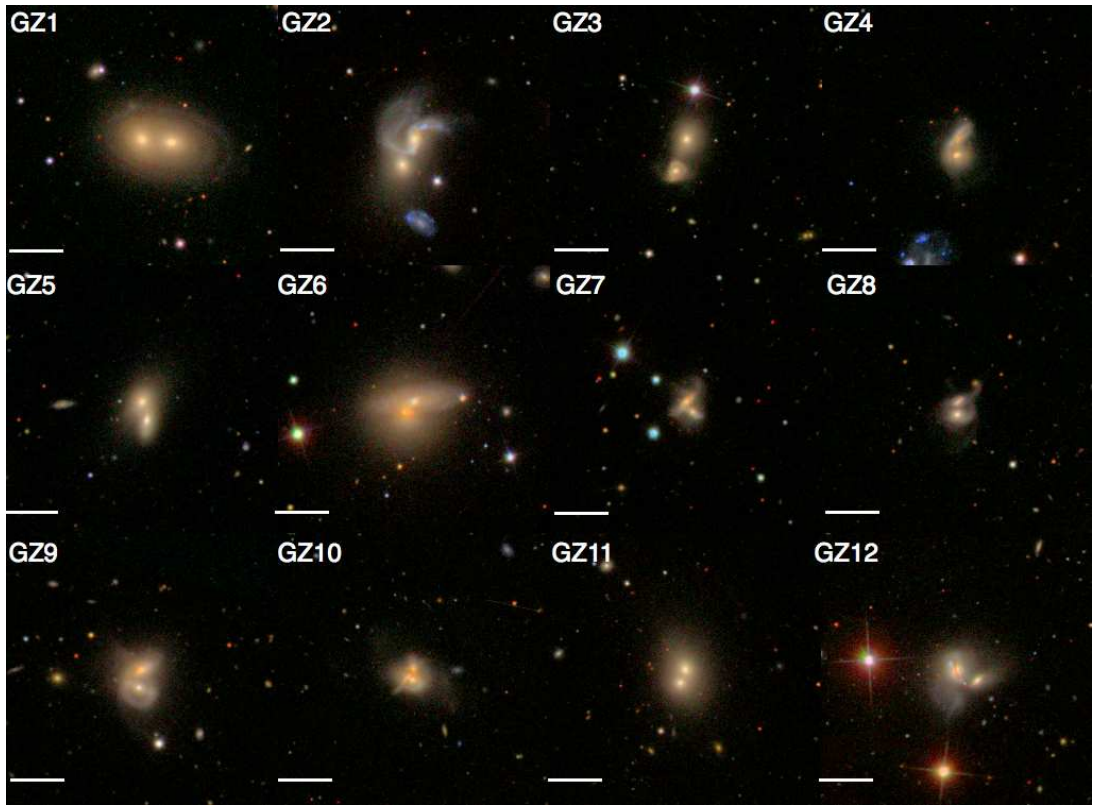}
\caption{Composite $gri$ images of the {\it Chandra}-observed sample from SDSS Data Release 7.  Each frame is labeled with the Galaxy Zoo identification listed in Table~\ref{tab:sample} and the horizontal bar represents angular distance of 20~arcseconds.}
\label{fig:sdss}
\end{figure}

The initial parent sample was created from 3003 mergers identified via visual inspection by citizen scientists taking part in the Galaxy Zoo project \citep{lintott, darg1, darg2}.  To date, this is the largest unbiased sample of mergers in the local Universe ($0.005 < z < 0.1$).  From this catalog, we whittled the sample down to only 328 major mergers (i.e., with a mass ratio of 3:1 or less) in which the primary galaxy has a mass\footnote{The stellar masses of the galaxies were calculated following the methodology outlined in \citet{schawinski10}.  Briefly, measurements from the five SDSS photometric bands were fitted to a library of model star formation histories generated from \citet{maraston98, maraston05} stellar models. Stellar masses are measured by finding the minimum of the $\chi^2$ statistic in the parameter space probed.} $> 10^{11} M_\odot$; we also required that the galaxies have SDSS spectroscopic data with signal-to-noise ratio greater than 3 and that at least one of the nuclei shows an AGN signature based on [O~III]/H$\beta$ and [N~II]/H$\alpha$ narrow line ratios \citep{bpt81, ke06}.  In addition, the AGN must be relatively luminous, with $L_{\rm [O\,III]} > 10^{41}$~erg~s$^{-1}$ (yellow points in Figure~\ref{fig:select}).  This last cut is to ensure that the AGN will be luminous enough for detection in the X-rays.  Most of the objects in this final sample are LINERs or AGN and star-forming composites, from which we selected 12 that spanned the full starburst-composite-LINER/AGN range.  The primary nuclei of the selected sample (yellow dots circled in red) also have roughly even distributions in the composite but not extreme starbursts (40\%) and the LINER (53\%) regimes of Figure~\ref{fig:select} and these mergers have projected nuclear separations between 5 to 14~kpc.  The optical line ratios are taken from \citet{bpt}.  The 12 galaxies in our sample are listed in Table~\ref{tab:sample}; for convenience, we will identify the targets as Galaxy Zoo (GZ) objects throughout this paper.   For clarity, we will refer to the merging galaxies as either ``mergers'' or ``galaxies'' and the individual progenitors of these mergers as ``nuclei'' for the remainder of this paper.  Figure~\ref{fig:sdss} is a collage of the SDSS images of the sample.  All of these mergers show disturbed morphology indicative of tidal disruptions.



\section{Observations, Data Reduction and Analysis}
\label{sec:obs}

The 12 mergers were observed with {\it Chandra} between November 2010 and May 2011 (PI: Schawinski).  Each pair of galaxies was observed in a single exposure of 4.9~ks.  For each merger, the more massive primary nucleus was placed at the nominal aim point of the ACIS-I3 chip.
The progenitors of the mergers are close enough that the secondary nuclei were also within the same chip.

The data were reduced using CIAO version 4.3 and CALDB version 4.4.3.  The data reduction followed procedures outlined in the Science Analysis Threads for ACIS imaging data on the CIAO web page\footnote{http://cxc.harvard.edu/ciao/threads/index.html}.  For each of the 12 detected nuclei, we extracted counts in the soft (0.5--2~keV) and hard (2--8~keV) bands.  Two of these nuclei were detected having only two counts in the soft band, but none in the hard band.  Therefore, only 10 nuclei have valid hardness ratios (HRs\footnote{$HR = \frac{H-S}{H+S}$, where H and S are the total counts in the hard and soft bands, respectively.}; Table~\ref{tab:sample}) for estimating spectral properties.  For the rest of this paper, only the 10 nuclei with valid HRs will be discussed.

An HR analysis was performed using the latest version of FTOOLS released as part of HEASoft 6.11.  Due to the low number of counts measured from our sample, we followed \citet{teng05} and used the measured HR to estimate a photon index ($\Gamma$) by assuming a redshifted power law model modified only by Galactic absorption.  The nominal HR and the estimated photon indices are tabulated in Table~\ref{tab:sample}.  The X-ray fluxes were then estimated using PIMMS by assuming the HR-derived photon indices and the count rates from the observations.  

\section{Discussion}
\label{sec:disc}

The shapes of the X-ray spectra differ for AGNs, obscured AGNs, starbursts, and AGN-star forming composites.  Typically, unobscured AGNs have spectra that are well-represented by a power law with photon index of $\sim$1.8.  Obscuration affects the lower energy ($\lesssim$2~keV) photons more readily than the higher energy photons and thus flatten or harden the AGN spectra.  Starburst spectra are dominated by emission in the lower energies, but low-mass X-ray binaries tend to have relatively flat spectra.  Composite objects generally have softened spectra compared to simple AGN spectra due to the soft-energy contribution of the starburst.  

Given that these mergers contain optically selected AGNs, it is unsurprising that eight of the 10 nuclei have HRs that are consistent with the canonical spectral shape of unobscured AGNs ($\Gamma \sim 1.7 - 2.1$).  As many as five could be steeper (GZ~1E, GZ~4S, GZ~5S, GZ~7SW, and GZ~9S), as if star formation is a significant contributor, though the errors in HR and $\Gamma$ allow for unobscured AGN values.   Similarly, three nuclei (GZ~2N, GZ~9N, and GZ~11S) have nominally flat spectra, implying dominance from star formation or obscured nuclear activity.   However, the errors in HR cannot rule out unobscured AGN as the source of the X-ray emission.  Finally, two nuclei (GZ~3S and GZ~10S) have flat or inverted photon indices ($\Gamma \lesssim 1.45$ after accounting for the measurement errors), suggesting some level of obscuration.  If we assume a power law with $\Gamma$ fixed at 1.8, the observed HRs imply column densities ($N_H$) $\lesssim 10^{21-22}$~cm$^{-2}$ (Table~\ref{tab:sample}).  At these column densities, the HR estimates of the 2--10~keV luminosity are reliable to within $\sim$40\% \citep{teng10}.  These columns do not suggest the presence of Compton-thick nuclei though there remains a possibility of leaky, Compton-thick absorbers.


\subsection{Starburst Contamination}

\begin{figure}
\centering
\plottwo{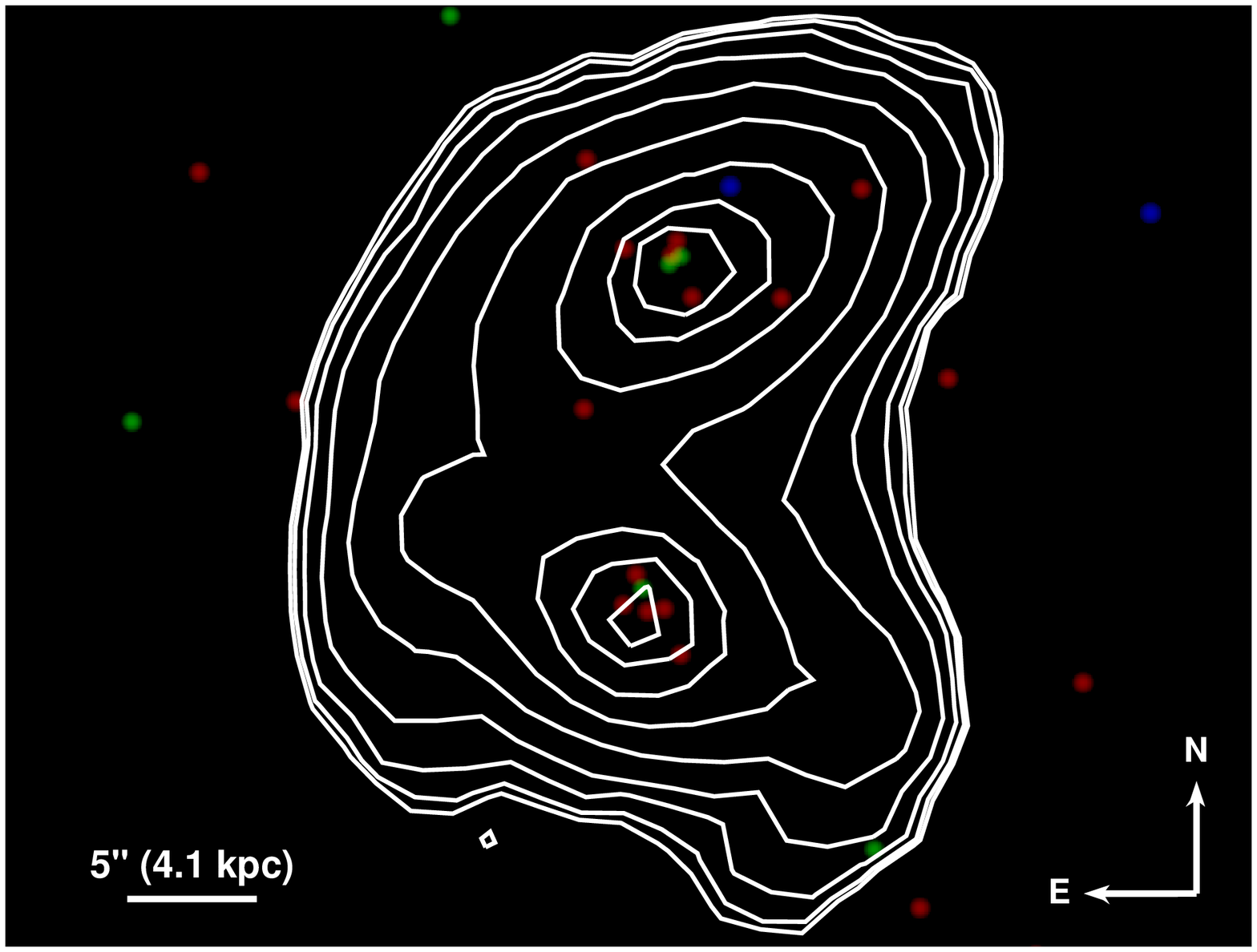}{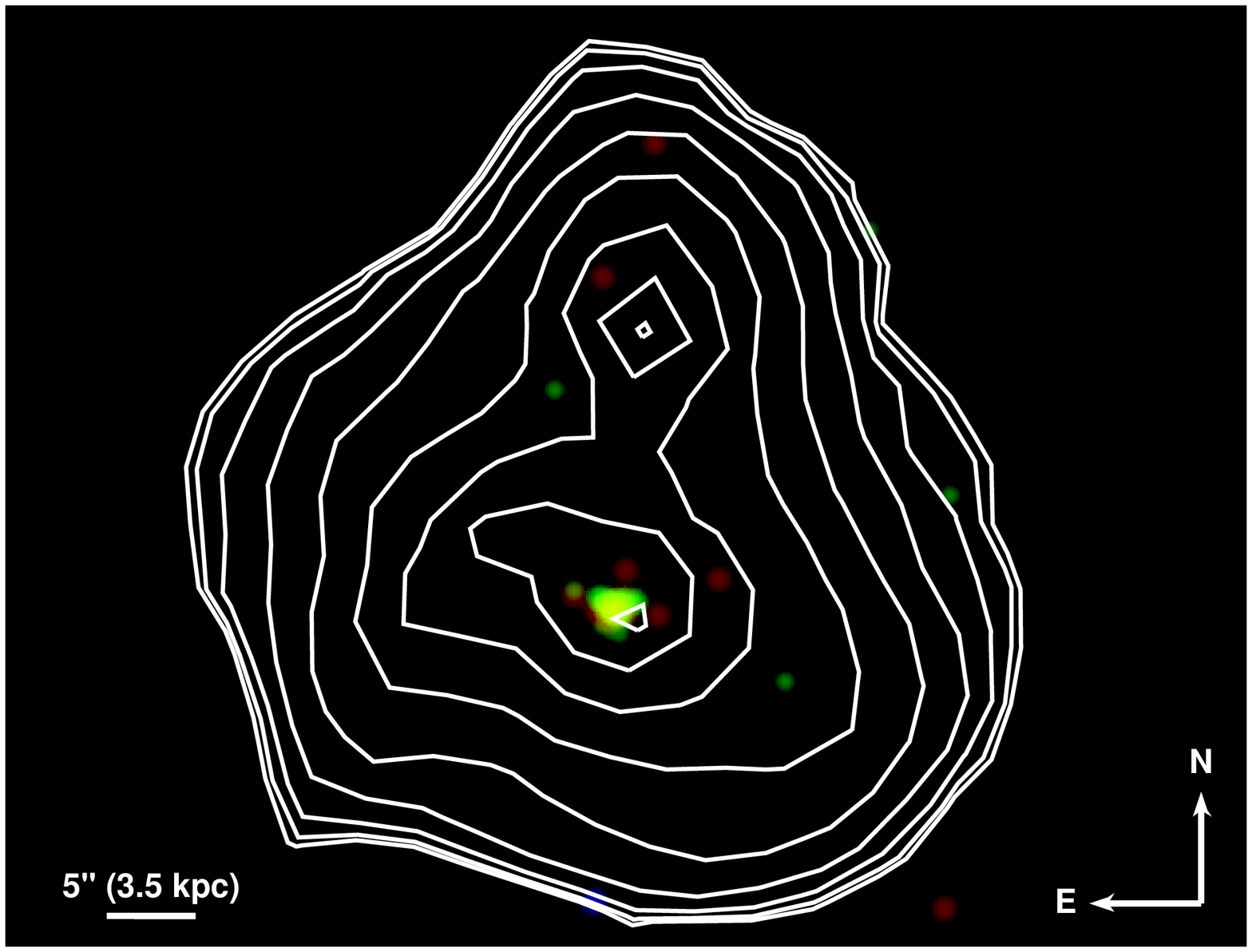}
\caption{False color X-ray images of two interesting objects in our sample, GZ~9 (left) and GZ~10 (right).  The red represent the 0.5--2~keV, green the 2--6~keV, and blue the 6--8~keV emission.  The contours are from SDSS {\it i} band images.   The raw X-ray images were smoothed with a 0\farcs5 Gaussian, the width of the nominal point spread function of {\it Chandra}.  The binary nuclei in GZ~9 are both detected in the X-ray, though the southern nucleus is dominated by soft X-ray emission.  In GZ~10, the X-ray emission shows east-west extension which may be due to star formation in addition to an obscured AGN.  The flatness of the X-ray spectrum implies a column density $\sim 5 \times 10^{21}$~cm$^{-2}$ if we assume a canonical power-law photon index of $\Gamma\,\sim\,1.8$. }
\label{fig:colorfig}
\end{figure}

Of the 12 mergers in the sample, one has no X-ray detection (GZ~8) and only one (GZ~9) exhibits binary X-ray nuclei (Figure~\ref{fig:colorfig}).  The remaining 10 mergers have one detected nucleus each.  In GZ~5, GZ~7, and GZ~12, the X-ray-detected nucleus is not the one with an optical AGN classification, so in that sense they are double nuclei.  In addition, the detected southern nucleus of GZ~10 has extended soft X-ray emission (Figure~\ref{fig:colorfig}), suggesting a contribution from star formation.  This raises the question of whether more of the detections might be contaminated by star formation.

\begin{figure}
\centering
\includegraphics[angle=270, scale=0.4]{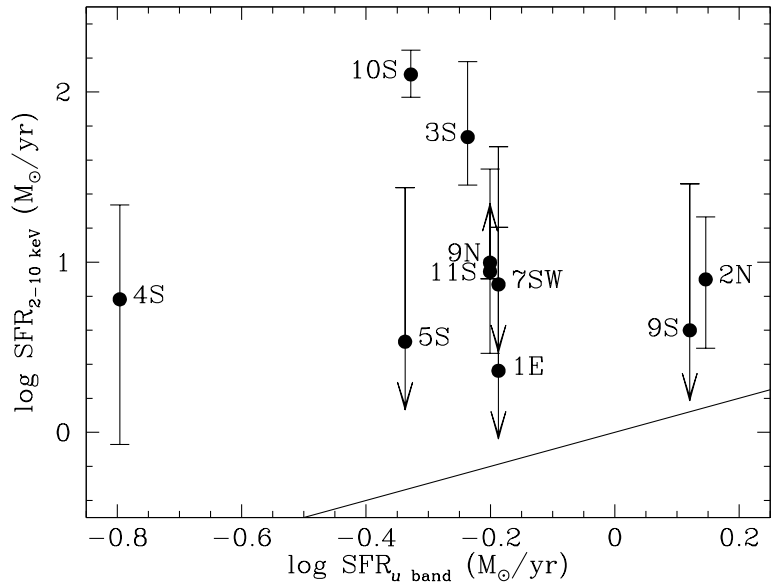}
\caption{Comparison of star formation rates derived from the 2--10~keV luminosity \citep{ranalli} and SDSS {\it u} band luminosity density \citep{sdss}.  Only the detected nuclei with hardness ratios, and thus estimated X-ray luminosities, are plotted.  Errors are $1-\sigma$.  The arrows indicate poorly constrained negative error bar for three nuclei whose HR lower limits approach --1, where the $N_H$ and $\Gamma$ values become degenerate for any value of HR.  The solid line is the line of equal star formation rates.  Each of the detected nuclei is labeled corresponding to the identification in Table~\ref{tab:sample}.  All of the detected nuclei have X-ray luminosity above that expected from star formation. }
\label{fig:sfrplot}
\end{figure}

To explore this possibility, we compare star formation rates derived from the SDSS {\it u} band luminosities following \citet{sdss} with those derived from the 2--10~keV luminosity following \citet{ranalli} in Figure~\ref{fig:sfrplot}.  When compared with the SDSS $u$-band derived star formation rates (Figure~\ref{fig:sfrplot}), the X-ray derived star formation rates of four nuclei (GZ~1E, GZ~5S, GZ~7SW, and GZ~9S) have unconstrained lower limits.  While the nominal X-ray derived star formation rates are above the line of equality implying the presence of AGNs, we cannot rule out the possibility that the X-ray emission can be accounted for solely by star formation in these four nuclei.  The X-ray luminosities of the remaining nuclei are above those expected from star formation even after the consideration of the $\lesssim$40\% error in the calculation of the X-ray luminosity, consistent with additional contribution to the X-ray luminosity by nuclear activity.  Accounting for the error bars, the southern nucleus in GZ~9 may also be dominated by star formation, suggesting GZ~9 does not contain an AGN pair.  

\subsection{Compton-thick Nuclei}

Three nuclei of the 12 SDSS-selected mergers are not detected in X-rays (GZ~5N, GZ~7NE, and GZ~12E).   
We already know these have optically identified AGN components, so it is unclear whether the non-detections are due to faint AGNs (two of the three have the highest redshifts in our sample) or Compton-thick AGNs.  If we assume these are faint AGNs, a power law model with $\Gamma$ = 1.8 and mild absorption from the Milky Way places upper limits to the luminosity of these objects.  Not accounting for intrinsic absorption, the 2-10~keV luminosity for GZ~5N is $\lesssim 5 \times 10^{39}$~erg~s$^{-1}$~cm$^{-2}$ and $\lesssim 5 \times 10^{40}$~erg~s$^{-1}$~cm$^{-2}$ for GZ~7NE and GZ~12E.  In the case of the Compton-thick AGNs, the optical signature is coming from the much larger scale narrow- and broad-line regions while the X-ray is sensitive to the small scale emission from the black hole itself.  The presence of undetected obscured nuclei would affect our statistics of the frequency of binary AGNs.  It is unlikely that all of the secondary nuclei contain Compton-thick X-ray sources, unless an obscured phase is common to mergers (unlike isolated AGNs).  Even without a merger-induced obscured phase, the number of heavily obscured AGNs is comparable to the number of less obscured AGNs \citep{tuv}; the presence of Compton-thick nuclei remains a possibility.  

\begin{figure}
\centering
\includegraphics[angle=270, scale=0.4]{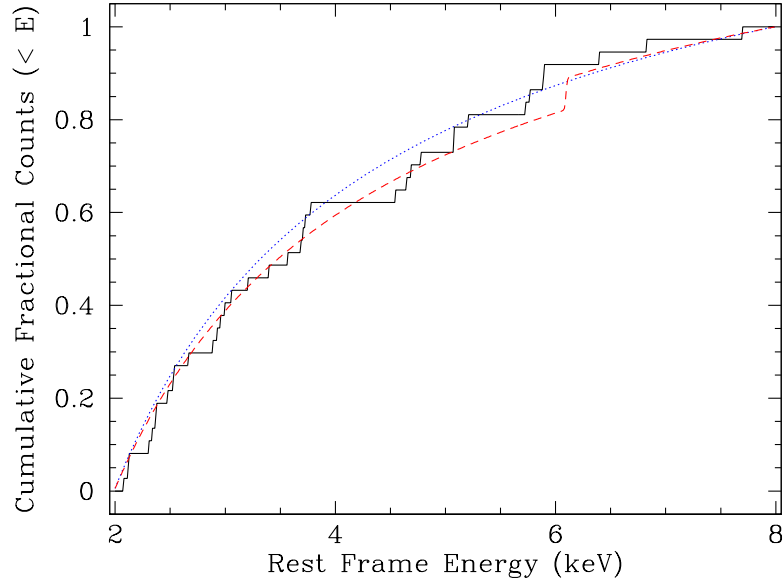}
\caption{Cumulative rest-frame photon distribution in the hard band (2--8~keV) for the 12 detected nuclei.  The black solid histogram represents the combined detected counts from all 12 detected sources in our sample.  The blue dotted curve represents the expected cumulative distribution assuming an unobscured AGN model where a single unabsorbed power law represents the source of the AGN emission.  The red dashed curve shows the distribution for a Compton-thick AGN model where the emission is represented by a power law plus a 6.4~keV iron emission line with an equivalent width of 1~keV.  While there is a clear difference between the two modeled distributions, when compared with the observed distribution, neither can be shown to be the preferred model by a two-tailed K-S test.  Therefore, we cannot rule out contribution from Compton-thick AGNs at present. }
\label{fig:cumuplot}
\end{figure}

While the individual detected nuclei have too few counts for spectral fitting to definitively establish whether Compton-thick AGNs are present, we considered the cumulative rest-frame photon distribution of the detected nuclei in the hard band.  We compared this observed distribution with the expected photon distributions from unobscured AGNs and from Compton-thick AGNs.  In the former case, we assumed a single unabsorbed power law; in the latter case, we assumed a power law with an iron emission line at 6.4~keV with an equivalent width of 1~keV, a typical signature of Compton-thick AGNs.  For both cases, the total photon counts were normalized to be the same as the total detected counts.   In Figure~\ref{fig:cumuplot} we plot the cumulative distribution of the detected photons in our sample.  There is no clear distinction between the observed distribution with either model.  In fact, the two-tailed Kolmogorov-Smirnov (K-S) test statistics for the two cases are nearly identical.  As a sanity check, we compared the two modeled distributions with each other and there is a clear difference at the $\sim$80\% confidence level.  Therefore, we cannot rule out the possibility that Compton-thick AGNs are present at the level that we are able to detected  these sources.

\subsection{U/LIRGs in Formation?}

In theoretical models of galaxy mergers \citep[e.g.,][]{hopkins08}, luminous and ultraluminous infrared galaxies (U/LIRGs) represent a stage that mergers go through before the formation of elliptical galaxies.  Initially, tidal torques enhance star formation and black hole accretion.  Then in the final coalescence of the galaxies, massive inflows of gas trigger starbursts with strengths similar to those inferred for U/LIRGs.  

The mergers in our sample appear to be the predecessors to U/LIRGs in this evolutionary picture.  The X-ray luminosities estimated for our mergers are approximately 10 times lower than those observed in most U/LIRGs, but are consistent with the lower end of the range measured in LIRGs \citep{teng10, lehmer, goals}.  This implies mergers in our sample are in the earliest stages of interaction, where the growth of the central black hole has not yet peaked.  

The incidence of binary AGNs in U/LIRGs is also rare.  The Revised Bright Galaxy Survey \citep[RBGS;][]{rbgs} is a flux-limited sample of U/LIRGs from the {\it IRAS} All Sky Survey.  Of the 629 extragalactic objects with 60~$\mu$m flux greater than 5.24~Jy, 86 are interacting galaxies that are visually similar to our sample in the optical (i.e. close binaries).  Of these, 32 have high-quality X-ray data from either {\it Chandra} or {\it XMM-Newton} that is sensitive to the presence of an AGN.  Not accounting for the presence of undetected Compton-thick nuclei, only 3\% (1/32) of the RBGS sources with X-ray data show binary X-ray nuclei \citep[NGC~6240;][]{komossa}.  This is consistent with the 0--8\% (0--1 out of 12) we observe in our modest SDSS sample.  

\subsection{Frequency of Binary AGNs in SDSS Mergers}

From the very short snapshots of our study, we have found that coincidence of optically selected active nucleus with mildly obscured X-ray nucleus is relatively common (8/12).  Given the faint detections, these snapshots are too short to place strong limits on the absence of AGN in the undetected galaxies, so it is difficult to comment on the frequency of binary active nuclei.  However, we do detect a pair of X-ray nuclei in GZ~9, implying that this is uncommon unless the second nucleus is heavily obscured.  In that instance, the most likely scenario would be that all nuclei are obscured.  That is, either binary nuclei are uncommon, or merger nuclei in general have a high probability of being heavily obscured.  The latter possibility cannot be addressed by the current sample.  To do better, we will need to increase the exposure times, expand our merger sample for better statistics, and include a sample of major mergers for which there are no optically detected nuclei.  Another natural follow-up would be to extend the study to a similarly selected sample with a lower mass limit to examine the dependence of binary AGNs on the mass of the host galaxies.



\acknowledgements

We thank the anonymous referee for his or her prompt and careful comments which improved the manuscript.  We made use of the NASA/IPAC Extragalactic Databased (NED), which is operated by the Jet Propulsion Laboratory, Caltech, under contract with NASA.  We acknowledge support by NASA through the \textit{Chandra} General Observer Program grant GO1-121401 to Yale University.  S.H.T. is supported by a NASA Postdoctoral Fellowship and K.S. is supported by NASA through an Einstein Postdoctoral Fellowship grant No. PF9-00069 issued by the Chandra X-ray Observatory Center, which is operated by the Smithsonian Astrophysical Observatory for and on behalf of NASA under contract NAS8-03060.

{\it Facilities:} \facility{{\it Chandra}}, \facility{SDSS}



\begin{turnpage}
\begin{deluxetable}{llcccccccccccccc}
\tablecolumns{16}
\tabletypesize{\tiny}
\setlength{\tabcolsep}{0.3cm}
\tablecaption{The Sample and Results}
\tablewidth{0pt}
\tablehead{
\colhead{AGN ID} &\colhead{SDSS ID}&\colhead{R.A.}&\colhead{Dec.}& \colhead{$z$}& \colhead{$N_{H, Gal}$}&\colhead{Sep.}&\colhead{Type}&\colhead{$\log$ M$_\star$}&\colhead{S}&\colhead{H}&\colhead{HR}&\colhead{Est. $\Gamma$}&\colhead{Est. $N_H$} &\colhead{$f_{2-10~keV}$} &\colhead{$L_{2-10~keV}$}\\
\colhead{(1)} & \colhead{(2)} & \colhead{(3)} & \colhead{(4)}& \colhead{(5)} & \colhead{(6)}&\colhead{(7)}&\colhead{(8)}&\colhead{(9)}&\colhead{(10)}&\colhead{(11)}&\colhead{(12)}&\colhead{(13)}&\colhead{(14)}&\colhead{(15)}&\colhead{(16)}
}
\startdata
GZ 1&&&&0.024&5.12&0.366& \\
\nodata E$^*$& 38195576881250&07:51:21.0&+50:14:10.0&&&(10.5)&L&11.19&3&1&	$-0.50^{+1.04}_{-0.50}$&2.12$^{+\infty}_{-2.17}$&$< 223.9$&0.46$^{+2.71}_{...}$&0.57$^{+3.40}_{...}$\\
\nodata W& 38195576881249&07:51:18.7&+50:14:08.0&&&&\nodata&11.18&\nodata&\nodata&\nodata&\nodata&\nodata&\nodata&\nodata\\
GZ 2&&&&0.026&1.22&0.352& \\
\nodata N$^*$& 39130806861890&14:01:41.4&+33:49:36.8&&&(10.9)&C&11.10&2&2	&$0.00^{+0.93}_{-0.46}$&1.00$^{+0.95}_{-2.94}$&	63.0$^{+731.3}_{...}$&	1.33$^{+1.77}_{-0.81}$& 1.96$^{+2.61}_{-1.19}$\\
\nodata S& 39130806861889&14:01:42.1&+33:49:17.6&&&&\nodata&10.89&\nodata&\nodata&\nodata&\nodata&\nodata&\nodata&\nodata\\
GZ 3 &&&&0.046&2.83 &0.416& \\
\nodata N& 36916218937372&15:11:20.9&+11:23:54.5&&&(22.3)&\nodata&11.18&\nodata&\nodata&\nodata&\nodata&\nodata&\nodata&\nodata\\
\nodata S$^*$& 36916218937373&15:11:21.5&+11:23:31.4&&&&L&11.05&3&4&$0.14^{+0.62}_{-0.36}$&	0.75$^{+0.70}_{-1.42}$&	95.0$^{+259.8}_{-5.9}$&	2.87$^{+5.12}_{-1.37}$&13.5$^{+24.0}_{-6.4}$\\
GZ 4&&&&0.028&2.67 &0.198& \\
\nodata N& 41532774793359&09:36:34.0&+23:26:39.3&&&(6.6)&S&10.68&\nodata&\nodata&\nodata&\nodata&\nodata&\nodata&\nodata\\
\nodata S$^*$& 41532774793358 &09:36:34.0&+23:26:27.0&&&&L&11.04&6&	2&$-0.50^{+0.62}_{-0.38}$&2.10$^{+1.68}_{-1.30}$&$< 89.1$&0.88$^{+2.26}_{-0.75}$&	1.50$^{+3.88}_{-1.29}$\\
GZ 5&&&&0.029&2.93 &0.233& \\
\nodata N$^*$& 29388212322361&08:46:20.2&+47:09:23.1&&&(8.0)&L&11.05&\nodata&\nodata&\nodata&\nodata&\nodata&\nodata&\nodata\\
\nodata S& 29388212322360&08:46:19.9&+47:09:09.3&&&&\nodata&11.01&3&1&$-0.50^{+1.04}_{-0.50}$&2.09$^{+\infty}_{-2.18}$&	$<$ 223.9&	0.46$^{+2.77}_{...}$&	0.85$^{+5.10}_{...}$\\
GZ 6&&&&0.029&0.92 &0.165& \\
\nodata NW& 29652348223597&16:29:57.5&+40:37:50.8&&&(5.7)&L&11.18&\nodata&\nodata&\nodata&\nodata&\nodata&\nodata&\nodata\\
\nodata SE$^*$& 29652348223595&16:29:58.1&+40:37:42.9&&&&L&11.66&3&0&$-1.0^{+2.0}_{-0.0}$&\nodata&\nodata&\nodata&\nodata\\
GZ 7&&&&0.048&4.03&0.148 & \\
\nodata NE$^*$& 34621631086789&08:38:17.9&+30:55:00.7&&&(8.2)&L&11.22&\nodata&\nodata&\nodata&\nodata&\nodata&\nodata&\nodata\\
\nodata SW& 34621631086790&08:38:17.6&+30:54:53.3&&&&\nodata&10.77&7&1&$-0.75^{+0.69}_{-0.25}$&2.95$^{+\infty}_{-1.80}$&$< 41.7$&0.36$^{+1.95}_{...}$&1.83$^{+9.99}_{...}$\\
GZ 8&&&&0.056&1.44 &0.161& \\
\nodata N& 38618094354457&10:22:56.5&+34:46:56.5&&&(10.4)&\nodata&10.82&\nodata&\nodata&\nodata&\nodata&\nodata&\nodata&\nodata\\
\nodata S$^*$& 38618094354456&10:22:56.6&+34:46:46.6&&&&L&11.10&\nodata&\nodata&\nodata&\nodata&\nodata&\nodata&\nodata\\
GZ 9&&&&0.033&1.18 &0.225& \\
\nodata N$^*$& 29156279631878&11:07:13.3&+65:06:06.5&&&(8.8)&L&11.14&4&2&$-0.33^{+0.72}_{-0.40}$&1.65$^{+1.15}_{-1.40}$&10.5$^{+151.7}_{...}$	&1.03$^{+2.63}_{-0.73}$&	2.46$^{+6.28}_{-1.74}$\\
\nodata S& 29156279631879&11:07:13.5&+65:05:53.2&&&&\nodata&10.81&5&1&$-0.67^{+0.82}_{-0.33}$&2.55$^{+\infty}_{-1.84}$&$< 89.1$&0.40$^{+2.18}_{...}$&	0.96$^{+5.21}_{...}$\\
GZ 10&&&&0.034&2.19 &0.123& \\
\nodata N& 39647816761431&10:47:11.2&+30:43:35.5&&&(4.9)&\nodata&11.00&\nodata&\nodata&\nodata&\nodata&\nodata&\nodata&\nodata\\
\nodata S$^*$& 39647816761430&10:47:11.2&+30:43:27.6&&&&L&11.08&20&19&$-0.03^{+0.20}_{-0.16}$	&1.06$^{+0.32}_{-0.37}$&	50.1$^{+52.2}_{-17.0}$&	12.4$^{+4.8}_{-3.3}$&	31.5$^{+12.2}_{-8.4}$\\
GZ 11&&&&0.039&1.34 &0.175& \\
\nodata N& 35661548929057&09:57:52.9&+36:20:57.5&&&(8.0)&\nodata&11.27&\nodata&\nodata&\nodata&\nodata&\nodata&\nodata&\nodata\\
\nodata S$^*$& 35661548929058&09:57:53.2&+36:20:47.1&&&&C&11.16&1&1&$0.00^{+1.00}_{-0.58}$&1.02$^{+2.23}_{-\infty}$&$63.1^{+\infty}_{...}$&0.65$^{...}_{-0.59}$&2.18$^{...}_{-1.98}$\\
GZ 12&&&&0.041&1.74 &0.281& \\
\nodata E$^*$& 38570323918904&13:52:26.7&+14:29:27.2&&&(13.5)&L&11.19&\nodata&\nodata&\nodata&\nodata&\nodata&\nodata&\nodata\\
\nodata W& 38570323918903&13:52:25.7&+14:29:19.3&&&&A&10.86&2&0&$-1.0^{+2.0}_{-0.0}$&\nodata&\nodata&\nodata&\nodata\\	
\enddata

\tablecomments{
Col.(1): Galaxy Zoo merger identifier in this paper.  $^*$ represents the nucleus in the pair with an optical spectrum from SDSS.  Col. (2): SDSS spectroscopic object ID with a prefix of  5877.  Col.(3)-(4): Right ascension and declination in J2000.  Col.(5): Redshift.  Col.(6):  Galactic column density in units of $10^{20}$~cm$^{-2}$\citep{nh}.  Col.(7): Projected optical separation of the nuclei in arcminutes (kpc). Col.(8) Optical spectral type classification based on Figure~\ref{fig:select} (A=AGN, L=LINER, C=AGN/Star-forming composite, and S = star-forming).  Col.(9): $\log$ of the galaxy mass, derived from the SDSS data \citep[e.g.,][]{schawinski10}.  Col.(10)-(11): Number of counts in the soft (0.5--2~keV) and hard (2--8~keV) bands, respectively.  Col.(12): Hardness ratio.  The  error bars are propagated assuming Poisson statistics \citep{cstat}.  Col.(13):  Photon index estimated from the observed hardness ratio assuming Galactic $N_H$.  Col.(14): Estimated intrinsic column density for the source assuming a power law with $\Gamma = 1.8$ in units of $10^{20}$~cm$^{-2}$.  Col.(15): Estimated 2--10~keV flux by assuming a redshifted power law with estimated $\Gamma$ from Column 13 normalized by the detected count rate.  The flux is given in units of $10^{-14}$~erg~~s$^{-1}$~cm$^{-2}$.  Col.(16): Estimated 2--10~keV luminosity based on flux in Column 15.  The luminosity is given in units of $10^{40}$~erg~s$^{-1}$.
}
\label{tab:sample}
\end{deluxetable}
\end{turnpage}


\end{document}